\documentclass[twocolumn,amsmath,amssymb,prl,superscriptaddress,longbibliography,nobalancelastpage]{revtex4-1}

\usepackage{titlesec}
\usepackage[colorlinks,bookmarks=false,citecolor=blue,linkcolor=blue,urlcolor=blue]{hyperref}
\usepackage{epsfig}
\usepackage{color}
\usepackage{mathtools}
\usepackage{subfigure}
\usepackage{multirow}
\usepackage{rotating}
\usepackage{lipsum}
\usepackage{graphicx} 
\usepackage{dcolumn}  

\usepackage{units}

\input{insbox}
\usepackage[titletoc,title]{appendix}
\usepackage[graphicx]{realboxes}
\usepackage{graphicx,subfigure}

        {\pagebreak[4]\global\pdfpageattr\expandafter{\the\pdfpageattr/Rotate 0}}%

\newcommand{\be}{\begin{equation}}
\newcommand{\ee}{\end{equation}}

\def\vec{\mathbf}
\def\mc{\mathcal}
\def\vec{\mathbf}

    \definecolor{treegreen}{RGB}{51, 102, 0}
    
    \definecolor{navy}{RGB}{0, 0, 128}

\begin{document}

\title{On the connection between magnetic interactions and the spin-wave gap of the insulating phase of NaOsO$_{3}$}

\author{Nikolaos Ntallis}
\affiliation{Department of Physics and Astronomy, Uppsala University, Uppsala 751 20, Sweden}

\author{Vladislav Borisov}
\affiliation{Department of Physics and Astronomy, Uppsala University, Uppsala 751 20, Sweden}

\author{Yaroslav O. Kvashnin}
\affiliation{Department of Physics and Astronomy, Uppsala University, Uppsala 751 20, Sweden}

\author{Danny Thonig}
\affiliation{School of Science and Technology, \"Orebro University, SE-701 82, \"Orebro, Sweden}

\author{Erik Sj\"oqvist}
\affiliation{Department of Physics and Astronomy, Uppsala University, Uppsala 751 20, Sweden}

\author{Anders Bergman}
\affiliation{Department of Physics and Astronomy, Uppsala University, Uppsala 751 20, Sweden}

\author{Anna Delin}
\affiliation{Department of Applied Physics, School of Engineering Sciences, KTH Royal Institute of Technology, AlbaNova University Center, SE-10691 Stockholm, Sweden}
\affiliation{Swedish e-Science Research Center (SeRC), KTH Royal Institute of Technology, SE-10044 Stockholm, Sweden}

\author{Olle Eriksson}
\affiliation{Department of Physics and Astronomy, Uppsala University, Uppsala 751 20, Sweden}
\affiliation{School of Science and Technology, \"Orebro University, SE-701 82, \"Orebro, Sweden}

\author{Manuel Pereiro}
\email[Corresponding author: ]{manuel.pereiro@physics.uu.se}
\affiliation{Department of Physics and Astronomy, Uppsala University, Uppsala 751 20, Sweden}

\begin{abstract}

The scenario of a metal-insulator transition driven by the onset of antiferromagnetic order in NaOsO$_3$ calls for a trustworthy derivation of the underlying effective spin Hamiltonian.
To determine the latter we rely on {\it ab initio} electronic-structure calculations, linear spin-wave theory, and comparison to experimental data of the corresponding magnon spectrum.
We arrive this way to Heisenberg couplings that are $\lesssim$45\% to$\lesssim$63\% smaller than values presently proposed in the literature and Dzyaloshinskii-Moriya interactions in the region of 15\% of the Heisenberg exchange $J$. These couplings together with the symmetric anisotropic exchange interaction and single-ion magnetocrystalline anisotropy successfully reproduce the magnon dispersion obtained by resonant inelastic X-ray scattering measurements. In particular, the spin-wave gap fully agrees with the measured one. We find that the spin-wave gap is defined from a subtle interplay between the single-ion anisotropy, the Dzyaloshinskii-Moriya exchange and  the symmetric anisotropic exchange interactions. The results reported here underpin the local-moment description of NaOsO$_3$, when it comes to analyzing the magnetic excitation spectra. Interestingly, this comes about from a microscopic theory that describes the electron system as Bloch states, adjusted to a mean-field solution to Hubbard-like interactions.  

\end{abstract}

\date\today
\maketitle


In the context of correlated electronic systems, 5$d$ oxides hold a distinct position, realizing the peculiar regime where Coulomb interactions (the Hubbard $U$ in particular), the $d$--$d$ hopping matrix elements (referred to as $t$'s), and the spin-orbit (SO) coupling ($\lambda$) have similar magnitudes ($\lambda\!\sim\!t\!\sim\!U$). In contrast, in 3$d$ and 4$f$ based magnets,  it is possible to identify a smallest energy scale, since the SO coupling defines a much smaller energy scale in 3$d$ transition-metal compounds ($\lambda\!\ll\!t\!<\!U$), and the 4$f$--4$f$ hopping is minute in rare-earth systems ($t\!\ll\!\lambda\!<\!U$).

The SO coupling in iridates and osmates, for example, is of order $\unit[0.5-0.6]{eV}$. In addition, the 5$d$ orbitals, being much more extended than, e.g., the rare-earth 4$f$ states, causes the 5$d$ electrons on neighboring ions to interact much more effectively. In other words, reasonably well localized magnetic moments can still be in place but ligand-mediated superexchange is greatly enhanced as compared to in, e.g., 4$f$ insulators. Combined with the strong SO coupling, it may generate highly anisotropic intersite magnetic interactions, with the remarkable situation encountered in $t_{2g}^5$ iridates of having either antisymmetric Dzyaloshinskii-Moriya (DM) \cite{Jackeli_2009,yadav} or symmetric Kitaev \cite{Jackeli_2009,Yamali_2014,Vamshi_NJP} effective coupling parameters that are even larger than the isotropic Heisenberg constant.

Since $t\!\sim\!U$, metal-insulator transitions (MIT's) may also occur. Extensively discussed in this respect is the osmium oxide perovskite compound NaOsO$_3$. However, rather than a Mott MIT, the scenario of a Slater MIT has been proposed for NaOsO$_3$ \cite{Shi_2009,Calder_2012,Du_2012,Jung_2013,Vecchio_2013}, where the formation of antiferromagnetic (AF) order is opening the insulating gap.

In this frame of reference, detailed knowledge of the underlying magnetic interactions is crucial.
Here we shed light on this matter by means of {\it ab initio} calculations and subsequent atomistic spin-dynamics simulations employing the derived effective magnetic interactions.
The computed magnon spectra are compared with experimental data as obtained by resonant inelastic X-ray spectroscopy (RIXS) \cite{Calder_2017}. We calculate from {\it ab initio} density functional theory (DFT) a Heisenberg exchange of $\sim \unit[4-6]{meV}$, that can be easily reconciled with the bandwidth of the magnon spectrum; $\sim$ 80 meV, reported on the basis of RIXS measurements\cite{Calder_2017}. However, our analysis shows that a single-ion anisotropy (SIA), the DM interaction, Heisenberg exchange as well as the symmetric anisotropic exchange interaction, are large, and necessary for theory to reproduce the magnon spectrum of NaOsO$_3$.


From the structural point of view, NaOsO$_3$ has the octahedral environment of Os$^{5+}$O$_6$, so that the electronic configuration is 5d$^3$, suggesting that the $t_{2g}$-band is half filled. Moreover, this compound shows Curie-Weiss metallic nature and abruptly but continuously turns to an AF insulator at $\unit[410]{K}$ \cite{Shi_2009}. 
The structural, electronic and magnetic properties of bulk NaOsO$_3$ are studied theoretically using density functional theory (DFT)\cite{Hohenberg1964,Kohn1965} within the generalized-gradient approximation in the PBE parameterization\cite{PBE1996}. Electronic properties are calculated within the all-electron full-potential fully relativistic approach, with linear muffin-tin orbitals as basis functions in both LDA+U and LSDA+U approximations, as implemented in the RSPt electronic structure code \cite{Wills1987,Wills2000,Wills2010}, with the U values ranged from 0 to $\unit[4]{eV}$.

\begin{figure}[b]
 \centering
 \vspace{-0.0cm}
 \subfigure{  \hspace{-0.5cm}
\raisebox{4.0cm}{\textbf{ a)}} \hspace{-0.25cm}
 \includegraphics[width=0.2\textwidth]{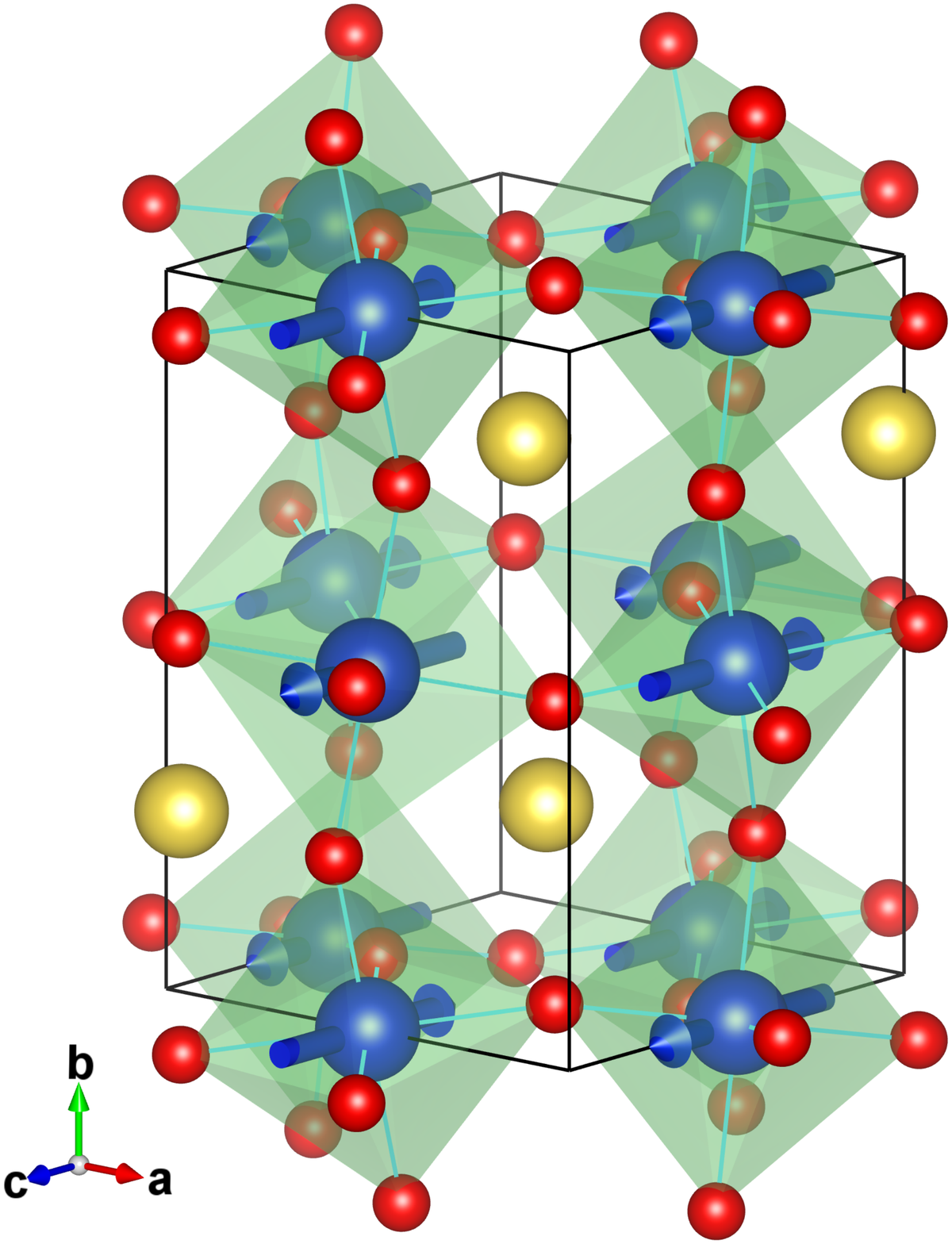}
\label{crystal_BaIrO3}}
 \quad
\subfigure{
\raisebox{4cm}{\textbf{ b)}} \hspace{-0.2cm}
 \includegraphics[width=0.185\textwidth,scale=1]{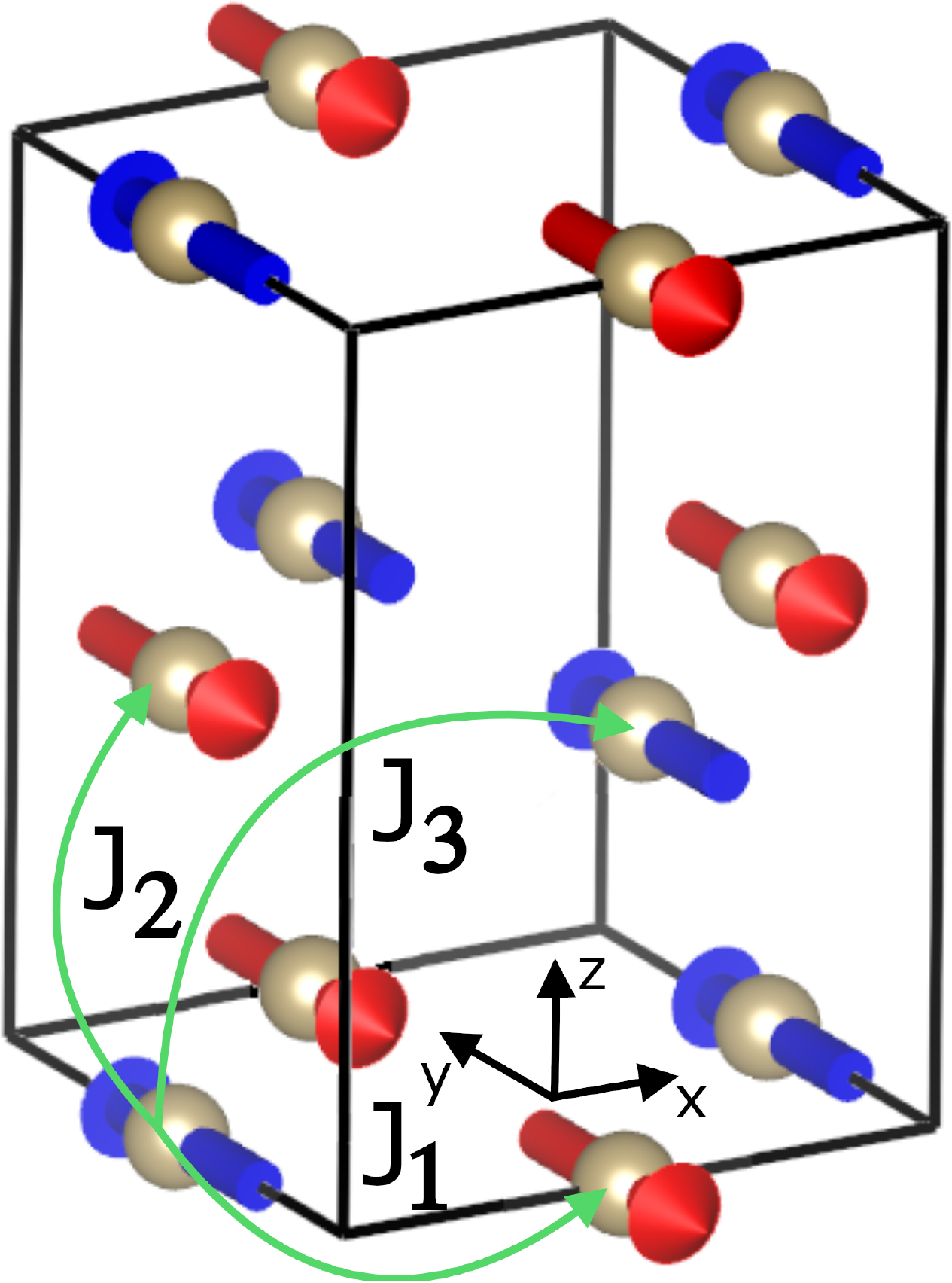}
\label{material_model}} \\
 \caption{
 a) $Pnma$ crystal structure of NaOsO$_{3}$. Red and yellow spheres indicate O and Na ions, respectively.
 b) $G$-type AF order within the Os sublattice. The NN exchange paths (within the $ac$ plane and along the $b$ axis) are highlighted.}
 \label{NaOsO3_str}
\end{figure}

\begin{figure}
\includegraphics[width=8.5cm]{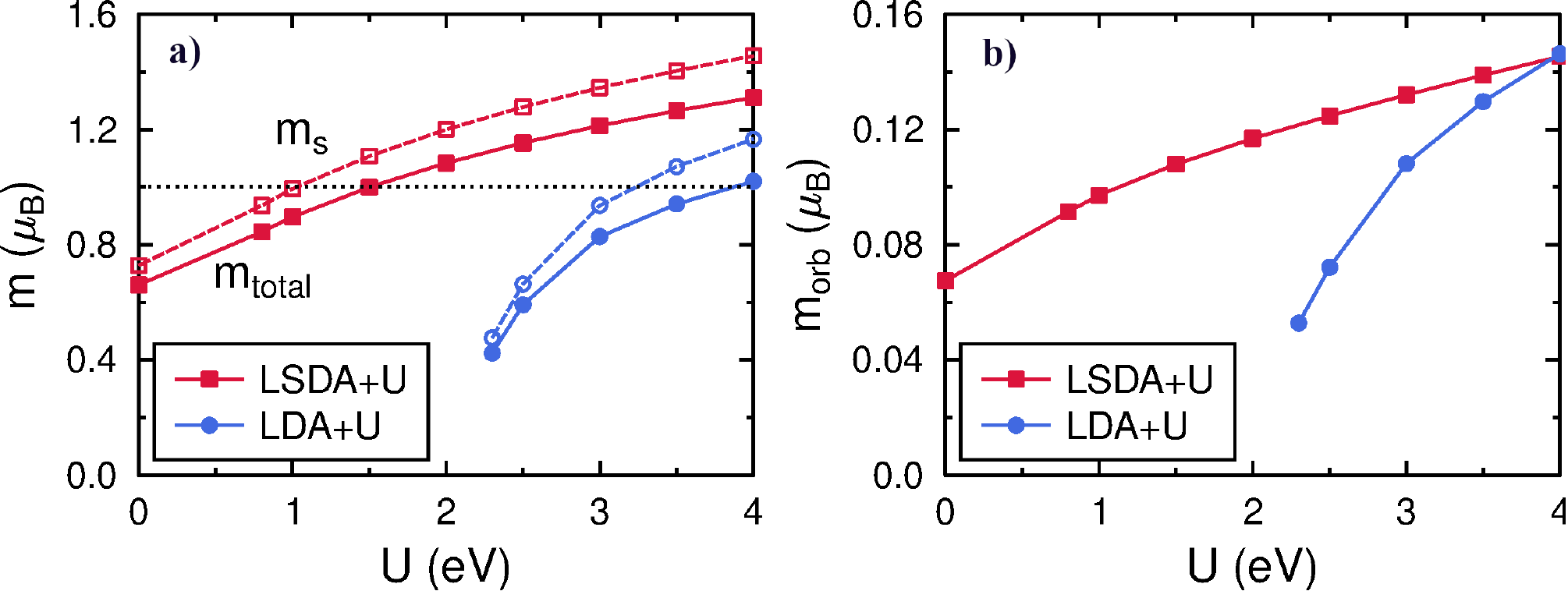}
\vspace{-10pt}
\caption{
Calculated total (solid symbols) and spin (unfilled symbols) magnetic moments in a) together with orbital moments (opposite to the spin moments) per atom of NaOsO$_{3}$ in b). The moments are plotted for different values of U and for LSDA+U and LDA+U implementations with fixed J$_{H}=0.6$ eV. The dotted line represents the magnetic moment measured in Ref.~\cite{Calder_2012}. 
}
\label{m_U}
\end{figure}

Figure~\ref{NaOsO3_str}(a) shows the experimental crystal structure of NaOsO$_3$. The cell dimensions are a=$5.3842$~\AA, b=$5.3282$~\AA~and c=$7.5804$~\AA. The magnetic ground state for all used values of U has a G-type antiferromagnetic (AF) ordering, as shown in Fig.~\ref{NaOsO3_str}(a). In Fig.~\ref{m_U}, the spin-, orbital- and total magnetic moment per Os atom is plotted with respect to the U value for the different approximations, i.e. LDA+U and LSDA+U. The two approaches produce different values of the magnetic moments as function of U. Notably, the LDA+U approximation would produce a zero magnetic moment for Os atoms by decreasing U down to 2 eV, according to Fig.~\ref{m_U}. This result is natural, since all exchange splitting in this level of approximation is in the static Hubbard type interaction, while the spin-density functional by construction provides no exchange splitting. In the limit of finite U, where both of the approximations produce almost the same orbital moment, the difference on the spin part is close to $\unit[0.35]{\mu_B}$ per atom. Also, one can reproduce the experimental values of the Os moment ($\sim$ 1 $\mu_B$) for both approximations, albeit for different values of U. 


Different types of magnetic interactions, such as the Heisenberg and DM interaction, were evaluated for the G-type antiferromagnetic reference state based on the fully-relativistic generalization \cite{Kvashnin2020} of the Lichtenstein-Katsnelson-Antropov-Gubanov (LKAG) formula \cite{LKAG1987}, which we have recently applied to different correlated systems \cite{Borisov2020}. 
The purpose of this is to map the electronic system onto a generalized classical Heisenberg model:
\begin{equation}
    H = -\sum\limits_{i\neq j} e_i^\alpha \hat J_{ij}^{\alpha\beta} e_j^\beta, \hspace{10pt} \alpha,\beta=x,y,z,
    \label{e:general_Heisenberg_model}
\end{equation}
where the unit vectors $\vec{e}_i$ indicate the direction of local spins and the fully-relativistic exchange tensor $J_{ij}^{\alpha\beta}$ contains contributions from the Heisenberg exchange as well as the DM interaction $\hat{D}$ and the symmetric anisotropic exchange $\hat{\Gamma}$ defined by
\begin{align}
    D_{ij}^z &= \frac12 (J_{ij}^{xy} - J_{ij}^{yx})\label{e:D_definition}\\
	\Gamma_{ij}^z &= \frac12 (J_{ij}^{xy} + J_{ij}^{yx})\label{e:C_definition}
\end{align}
and similar expressions for the $x$- and $y$-components. 

\begin{figure}
\includegraphics[width=8.5cm]{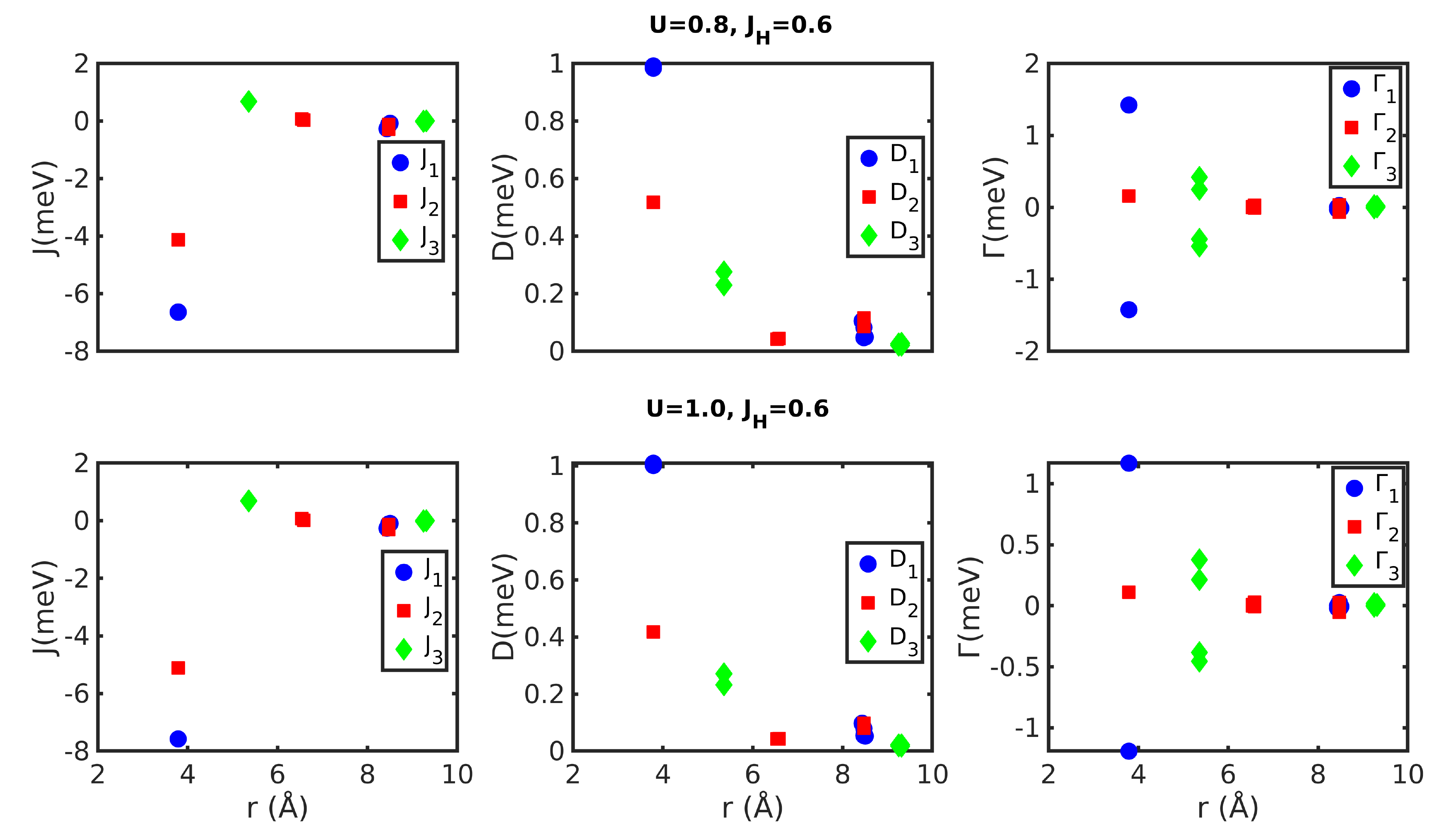}
\caption{
Calculated interaction strengths $J+D+\Gamma$ with respect to the atomic distance for U=0.8 eV, J$_{H}=0.6$ eV (top panel) and U=1.0 eV, J$_{H}=0.6$ eV (bottom panel). All interaction are marked following the prescription  $1\rightarrow 3 $ given in Fig.~\ref{NaOsO3_str}.
 }
 \label{nei_J_D}
\end{figure}

According to Fig.~\ref{m_U}, the values of U which produce total magnetic moments  closest to the experimental ones are U $\approx \unit[0.8]{eV}$ for LSDA+U, similarly to \cite{Middey2014,Mohapatra2018,Liu2020}, and U $\approx \unit[3.5]{eV}$ for LDA+U.
Figure ~\ref{nei_J_D} shows the calculated magnetic interaction strengths with respect to the atomic distance. All interaction strengths decay quite fast with respect to distance indicating that the major interactions to be considered are the first and second order ones. The LDA+U approximation (with U = $\unit[3.5]{eV}$) produces Heisenberg exchange interaction that is roughly twice the magnitude compared to the LSDA+U data (with U = $\unit[0.8]{eV}$). Similarly the LDA+U calculation results in a value of $D$ that is approximately three or four times larger than the one for LSDA+U approximation. In order to assess the quality of the exchange interactions we used them to evaluate adiabatic magnon spectra, that can be compared to experimental results (see discussion below). We note, however, already here that such an analysis reproduces experimental results quite poorly from a theory based on LDA+U (see supplementary information (SM) for further details on the LDA+U results)\cite{supmat}, while LSDA+U calculations give reasonable results. For this reason we focus in the rest of this paper on results from LSDA+U calculations.


\begin{figure}
\includegraphics[width=0.99\columnwidth]{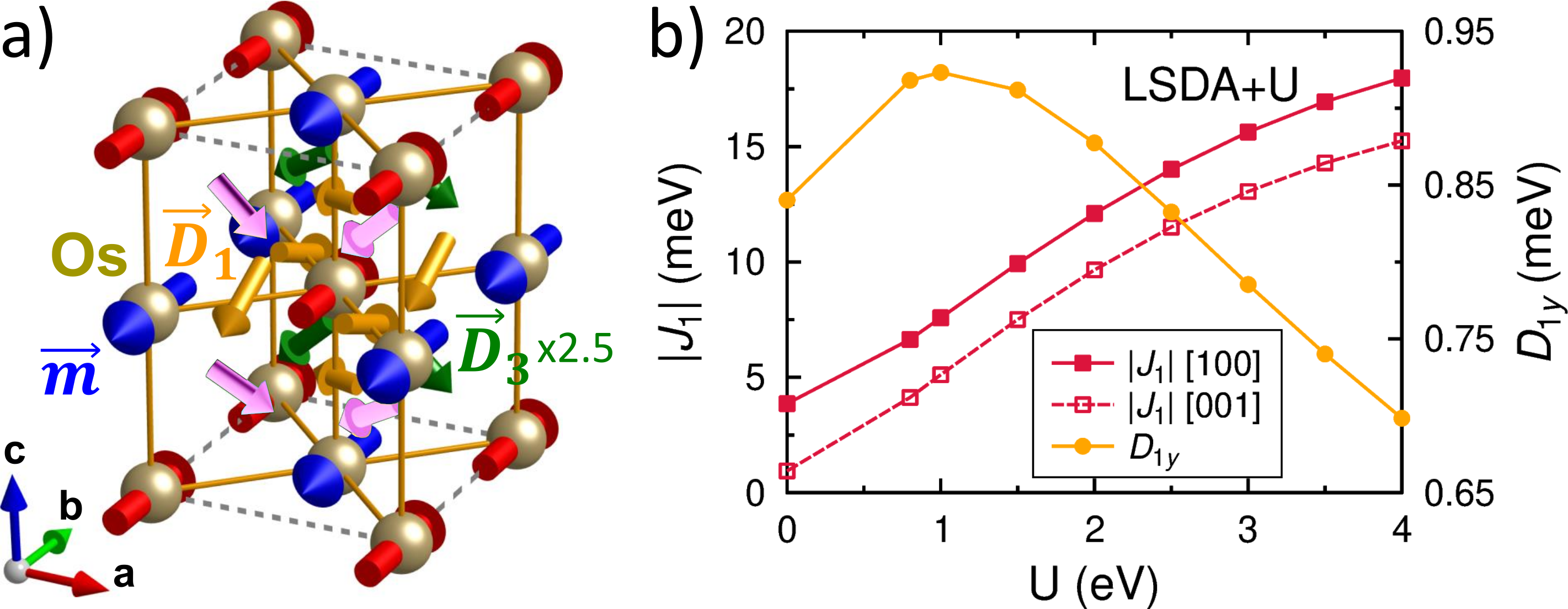}
\caption{
a) Calculated DM vectors $\vec{D}$ for the nearest neighbor bonds in NaOsO$_3$ shown by the orange ($\vec{D}_1$), green ($\vec{D}^a_3$) and magenta ($\vec{D}^b_3$) arrows. The $\vec{D}^{a,b}_3$ vectors are magnified by the factor of 2.5 for clarity. The G-type antiferromagnetic ordering of Os magnetic moments $\vec{m}$ oriented along the $y$-direction is depicted by the red and blue arrows. The magnetic interactions were obtained using LSDA+$U$ with $U = \unit[0.8]{eV}$ and $J_{H} = \unit[0.6]{eV}$. b) The nearest-neighbor Heisenberg exchange $J_1$ along the [100] and [001] directions and the $y$-component of the DM interaction $\vec{D}_1$ in the \textit{ab} plane as functions of the correlation strength $U$.
 }
 \label{f:D_vectors}
\end{figure}


Despite the same magnetic pattern along the $x$ and $z$ directions and very similar distances ($d_x=\unit[3.787]{A}$ and $d_z=\unit[3.790]{A}$), the Heisenberg exchange interactions $J_{1x}$ and $J_{1z}$ are markedly different. This could be related to the large effect of magnetic anisotropy which aligns the magnetic moments and the N\'eel vector preferably along the $y$ direction. It is interesting that the $J_{1z}/J_{1x}$ ratio increases as a function of the correlation strength U from 0.62 at $\mathrm{U}=\unit[0.8]{eV}$ to 0.85 at $\mathrm{U}=\unit[4]{eV}$. This may suggest that the system becomes magnetically more isotropic for stronger correlations where the system is more insulating, based on the calculated value of the electronic band gap. We also notice that the scaled Heisenberg parameters $J/m_s^2$ are far from the $\nicefrac{1}{U}$ scaling expected for insulators. This can be explained by the fact that correlations in the narrow-gap insulator NaOsO$_3$ are not strong compared to the electron hopping amplitude $t$, so that $\nicefrac{t}{U}$ is not a small parameter and the perturbation theory, which could lead to the $\nicefrac{1}{U}$ scaling, is not applicable. Interestingly, the nearest-neighbor parameters $J_1/m_s^2$ increase as functions of U up to $\mathrm{U}=\unit[3.0-3.5]{eV}$ which deserves a more detailed study in the future. On the other hand, the DM interaction shows a non-monotonous variation of its $y$-component, which is dominating for the bonds along the $xy$-directions, with a maximum around $\mathrm{U}=\unit[1]{eV}$ followed by a decreasing trend for larger U values.

The numerical values of the exchange interactions are collected in Table~\ref{Os_DM} for LSDA+U calculations, while the DM vectors for the nearest-neighbor bonds are shown in Fig.~\ref{f:D_vectors}. These interactions were used to calculate the magnon spectra, which was evaluated by rewriting Eq.~\eqref{e:general_Heisenberg_model}, in the form of  
a bilinear, effective Hamiltonian
$\mc{H}_{i,j}$ containing the isotropic Heisenberg, the DM as well as symmetric anisotropic exchange.
Adding single-site magnetic anisotropy we end up with the following expression for the full Hamiltonian of the spin system:

\begin{eqnarray}
\label{eq_spin}
    \mc{H}^{i,j}_\mathrm{mod}
        &=& -J_{ij} \vec{{S}}_i \cdot \vec{{S}}_j
         - \vec{D}_{ij} \cdot \vec{{S}}_i\times \vec{{S}}_j 
         -\vec{{S}}_i \Gamma_{ij} \vec{{S}}_j\\ 
        &-& K \sum_{k=i,j}
               \left(\vec{S}_k \cdot \vec{e}^\mathrm{Y}_{k}\right)^2,\nonumber
\end{eqnarray}
where $\vec{{S}}_i$, $\vec{{S}}_j$ are normalized atomic spin moments. The isotropic Heisenberg coupling is given by $J_{ij}$ and $\vec{D}_{ij}$ stands for the DM vector, while the tensor $\Gamma_{ij}$ represents the symmetric anisotropic exchange interaction between atomic sites $i$ and $j$. The Heisenberg exchange interaction $J_{ij}$, the norm of $\vec{D}_{ij}$ and mean value of $\Gamma_{ij}$  tensor take the values $J_1$, $\left | \vec{D}_1 \right |$, $\Gamma_1$ and $J_2$, $\left | \vec{D}_2 \right |$, $\Gamma_2$ for bonds 1 and 2 (cf. Fig.~\ref{NaOsO3_str}(b)), respectively. 
The orientation of the DM vectors depend on the position in the unit cell, see Fig.~\ref{f:D_vectors}. The DM vectors $\vec{D}^a_3$ and $\vec{D}^b_3$ for the third-neighbours atoms are slightly different (cf. Table~\ref{Os_DM} and Fig.~\ref{f:D_vectors}) because the cell dimensions in the ab plane are also slightly different (a$\ne$b). As regards the magnetic anisotropy, each site possesses an easy-axis orientation directed along the local
magnetic y-axis $\vec{e}^{\mathrm{Y}}_{k}$. The magnetic anisotropy parameter ($K$ in Eq.~\eqref{eq_spin}) was calculated based on the sum of eigenvalues in the one-shot relativistic calculations using the RSPt code \cite{Wills1987,Wills2000,Wills2010} where the magnetization axis is oriented along the $x$-, $y$- and $z$-directions. The obtained estimates are $K=\unit[2.05]{meV}$ for U=$\unit[0.8]{eV}$ and $J_{H}=\unit[0.6]{eV}$ while the calculation with U=$\unit[1.0]{eV}$ and $J_{H}=\unit[0.6]{eV}$ resulted in $K=\unit[2.44]{meV}$.


\begin{table}[t]
 \caption{Intersite interaction vectors, effective couplings, and DM vectors for different Os-O-Os bond angles. The units of the coupling parameters are given in meV. The reference system is specified in Fig.~\ref{NaOsO3_str}(b).}
\vspace{5pt}
 \begin{ruledtabular}
 \begin{tabular}{lcc}
    $\theta_1=155.2^{\circ}$ & $U=\unit[0.8]{eV}$ &   $U=\unit[1.0]{eV}$   \\
    $J_1$     &    -6.64   & -7.58 \\
    $\Gamma_1$&     0.62   &  0.57   \\
    $\vec{r}_{ij}$  & $ \left (\pm 0.5, \pm 0.5,  0 \right )$& $ \left (\pm 0.5, \pm 0.5,  0 \right )$\\
    $\vec{D}_1$  & $\pm \left ( 0.04, -0.92, -0.37
\right )$ & $\pm \left ( 0.08,-0.92 ,-0.40 \right )$ \\
    \hline
    $\theta_2=153.9^{\circ}$ &  &   \\
    $J_2$     &    -4.12   & -5.11 \\
    $\Gamma_2$&     0.07   &  0.05 \\
    $\vec{r}_{ij}$  & $ \left (0.0, 0.0,\pm 0.5 \right )$& $ \left (0.0, 0.0,\pm 0.5 \right )$\\
    $\vec{D}_2$ & $\left ( -0.50, 0.15, 0.00
\right )$ & $\left ( -0.41, 0.07, 0.00\right )$ \\
    \hline
    $\theta_3$=$\emptyset$ &  &   \\
    $J_3$     &    0.67   & 0.68 \\
    $\Gamma_3$&     0.18   &  0.18   \\
    $\vec{r}_{ij}$  & $ \left (\pm 0.5, \pm 0.5,\pm 0.5 \right )$& $ \left (\pm 0.5, \pm 0.5,\pm 0.5 \right )$\\
    $\vec{D}^a_3$  & $\pm \left ( 0.20, -0.01, -0.18 \right )$ & $\pm \left ( 0.20, -0.03,-0.18 \right )$ \\
$\vec{D}^b_3$ & $\pm \left ( -0.20, 0.01, -0.11 \right )$ & $\pm \left ( 0.20, 0.00,-0.10 \right )$ \\

 \end{tabular}
 \end{ruledtabular}
    \label{Os_DM}
\end{table}



\begin{figure}
\includegraphics[width=8.5cm]{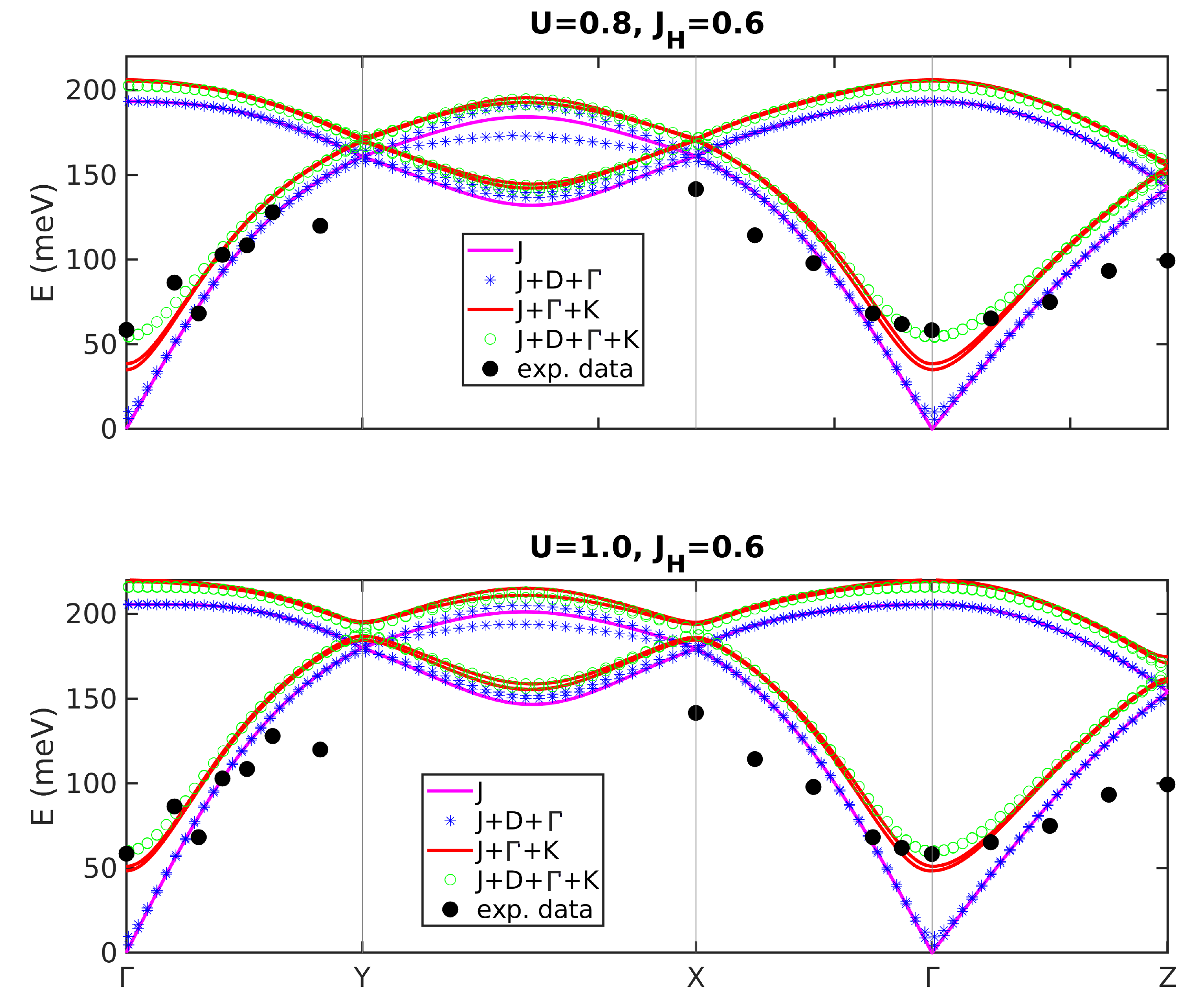}
\caption{
Adiabatic magnon spectrum of NaOsO$_3$  by progressively adding energy contributions to the effective spin Hamiltonian.
We illustrate  the cases with contributions
given by $J$, $J\!+\!\vec{D} + \!\vec{\Gamma}$, $J\!+\!\vec{\Gamma} +\!K$, $J\!+\!\vec{D} + \!\vec{\Gamma} +\!K$, for U=0.8 eV, J$_{H}=0.6$ eV from LSDA+U theory (top panel) and U=1.0 eV, J$_{H}=0.6$ eV from LSDA+U theory  (bottom panel). Experimental data provided by RIXS measurements are shown in black dots \cite{Calder_2017}.
}
\label{fig_ams}
\end{figure}


In order to validate the parameters calculated by {\it ab initio} electronic structure theory, we performed calculations of magnon dispersions, based on a collinear AF ground state, in the framework of linear spin-wave theory (LSWT) \cite{yadav} by using the UppASD software \cite{asd}. Effective coupling parameters as collected in Table~\ref{Os_DM} were utilized, along with a moment as illustrated in Fig.~\ref{m_U}a), from LSDA+U approximation using U=0.8-1.0 eV and J$_H$= 0.6 eV.

In order to understand which interactions dominate the spin-wave gap of NaOsO$_3$, we calculated adiabatic magnon spectra by progressively adding different energy terms as present in Eq.~(\ref{eq_spin}), see Fig.~\ref{fig_ams}. It is evident from the figure that the uniaxial single ion anisotropy is the term responsible for the spin-wave gap and the value of U that best fit the experimental spin-wave gap is achieved by U=1.0 eV while for U=0.8 eV, the gap is slightly too low (the magnon excitations are evaluated for a wider range of values of U in the SM\cite{supmat}). 
Furthermore, the contribution of the DM interaction together with the symmetric anisotropic exchange interaction even though they are small, they are necessary to correctly describe the full experimental spin-wave dispersion. Interestingly, the calculated profile of the magnon spectra for U=1 eV agrees quite well with the experimental spectra, even though the exchange coupling parameters together with the anisotropy constant differ substantially with respect to the ones suggested from the experimental data in Ref.~\cite{Calder_2017}. A direct fitting procedure of the magnon spectra, with a relative large number of parameters in a spin-Hamiltonian, can agree with experimental data but the parameters of the Hamiltonian obtained from such fittings, might be completely off compared to interactions calculated from {\it ab initio} electronic structure theory. This illustrates that {\it ab initio} theory is crucial for extracting proper interaction types, strengths and ranges.

\begin{figure}[b]
\includegraphics[width=8.7cm]{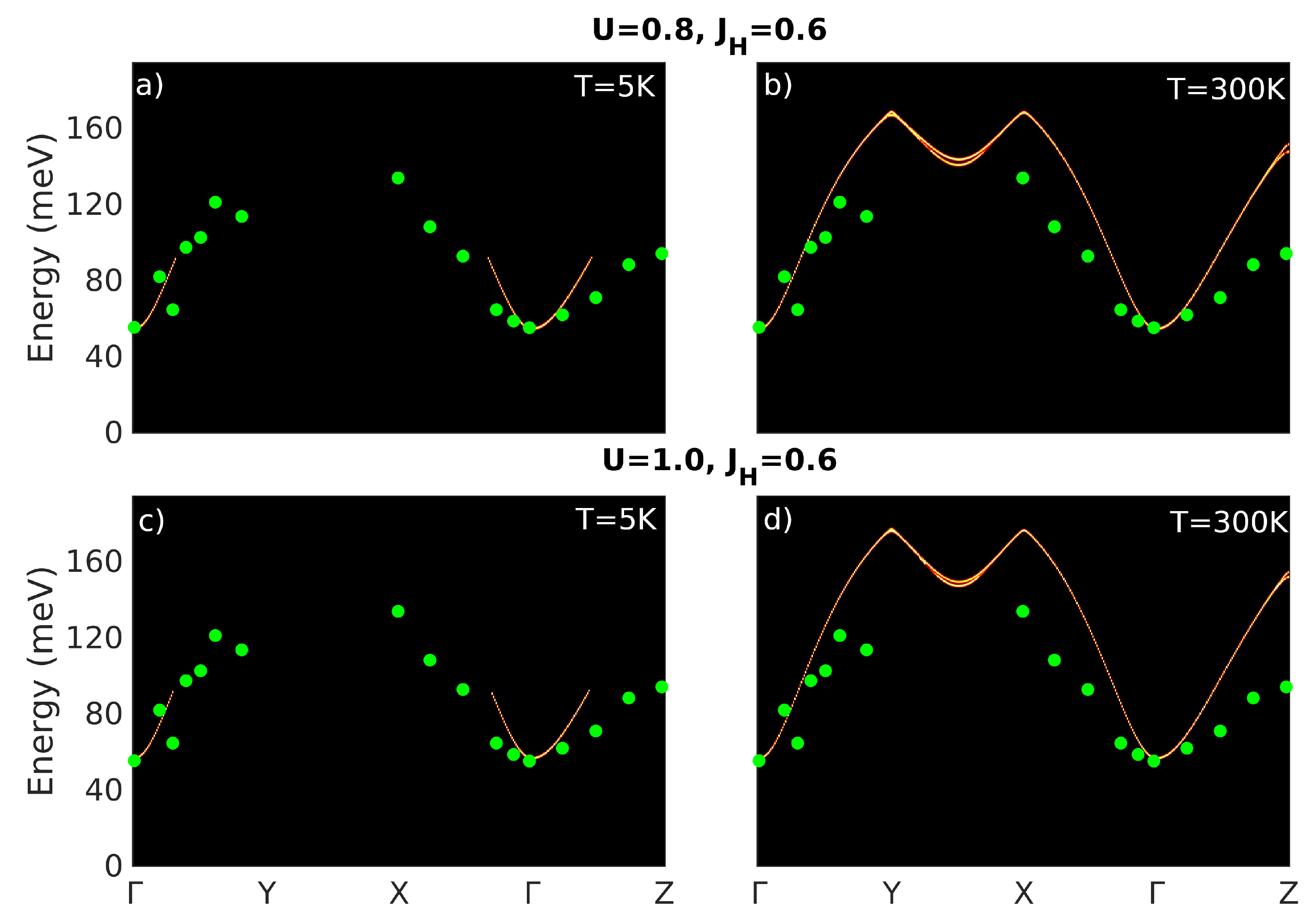}
\caption{
Dynamical structure factor of NaOsO$_3$ for a) U=0.8 eV, J$_{H}=0.6$ eV  and at T=5K, b)  for U=0.8 eV, J$_{H}=0.6$ eV  and at T=300K, c) for U=1.0 eV, J$_{H}=0.6$ eV  and at T=5K and d) U=1.0 eV, J$_{H}=0.6$ eV  and at T=300K. All the aforementioned parameters are calculated at the level of LSDA+U theory. Energy is convoluted  with a 0.7 meV full width at half maximum Gaussian. Experimental data provided by RIXS measurements are shown by green dots \cite{Calder_2017}.
}
\label{fig_dcf}
\end{figure}

The calculated adiabatic magnon excitation energies are in good agreement with the experimental data reported in Ref.~\cite{Calder_2017}, as shown in Fig.~\ref{fig_ams}. However, AMS is formally calculated at T$=0$K, whereas the experimental measurements are done at room temperature, i.e. T$=300$K. Therefore theory and experiments refer to data obtained at different temperatures. 
In order to shed more light on this point, we show in Fig.~\ref{fig_dcf} the dynamical structure factor  calculated in the framework of LSWT, which is proportional to the differential cross section measured in scattering experiments. It depends on the crystal momentum ($\bf q$) and energy transfer ($\omega$) and is defined as a 3$\times$3 matrix:
\begin{equation}
	S({\bf q},\omega)=\frac{1}{2 \pi N}\sum_{i,j}e^{i{\bf q}({\bf r}_{i}-{\bf r}_{j})}\int_{-\infty}^\infty e^{-i\omega t} \langle {\bf S}_{i}(t){\bf S}^T_{j}(t)\rangle dt\,,
\end{equation}
where indices $i, j$ label the atomic positions and ${\bf r}_{i}$ is the position vector of atom $i$.
$N$ is the number of magnetic atoms in the unit cell and ${\bf S}_{i}$ represents a column vector of the $x$, $y$, and $z$ spin operator components.
The symbol $\langle ...\rangle$ denotes the ensemble average in thermal equilibrium with the environment at temperature $T$.
Since we are dealing with magnons, i.\,e., bosons,
$T$ is introduced in the thermodynamic average by using Bose-Einstein statistics, with a Bose factor
\begin{equation}
n(\omega)=\frac{1}{e^{\hbar \omega/k_B T}-1} \,.
\end{equation}

In  Fig.~\ref{fig_dcf}(a),(c), we show $S({\bf q},\omega)$ for NaOsO$_3$  using magnetic interactions according to Eq.~(\ref{eq_spin}) and Table~\ref{Os_DM} at $T\!=\!5$ K and in Fig.~\ref{fig_dcf}(b),(d) the dynamical structure factor is calculated for $T\!=\!300$ K, the value of the temperature at which the experimental measurements have been reported in Ref.~\cite{Calder_2017}.
The dispersion of the curve in Fig.~\ref{fig_dcf} is naturally the same as in Fig.~\ref{fig_ams} but additional information is obtained from the intensity, which reveals how likely a scattering event is. Notably, for $T\!=\!5$ K only in the vicinity of the $\Gamma$ point does the cross section show a significant intensity indicating that at very low temperatures only these specific modes would be captured. On the other hand, at a temperature of $T\!=\!300$ K, the two lowest energy branches  show a considerable intensity along the $\bf q$-path in agreement with experimental data. Note here that according to Fig.~\ref{fig_ams}, the first acoustic and  first optical branches are very close in energy and thus both of them are activated by increasing temperature. Figure~\ref{fig_dcf} also reveals that at 300 K, the experimental conditions of the RIXS measurements, the experimental apparatus only can capture the acoustic and first optical branches while the additional  third and fourth optical branches remain hidden to the experiment. 

In summary, the very rare situation in which {\it ab initio} calculations of Heisenberg exchange results in $\lesssim$45\% to $\lesssim$63\%  smaller values compared to those obtained through fitting of experimental magnon dispersions by LSWT, has led us to a careful analysis of the spin excitation spectrum of the osmate perovskite NaOsO$_3$. To this end, we carried out atomistic spin-dynamics simulations using as input the effective magnetic interactions determined from {\it ab initio} calculations. These interactions involved not only Heisenberg but also DM and symmetric anisotropic intersite couplings, plus single-ion spin anisotropies. The effective DM interaction parameters turn out to be sizable, $\sim$14\% of the Heisenberg $J_1$'s for U=$\unit[1]{eV}$, while the single-ion anisotropies are also robust. Our theoretical results reproduce with good accuracy the observations, albeit with a distinct difference in that the spectra along the X $\rightarrow\Gamma\rightarrow$ Z path are slightly steeper than the measured spectra. Moreover, the third and fourth optical branches demonstrated in the theory, are not found in the experimental measurements. We propose a mechanism based on thermal fluctuations for why these two modes are hidden from experimental detection. The analysis put forth here differs from that of Ref.~\cite{Calder_2017} in that our calculations provide somewhat smaller magnetic anisotropy, but notably, a sizable DM exchange which was not suggested in Ref.~\cite{Calder_2017}. We note that the DM interaction can on general grounds not be neglected in NaOsO$_3$ since the O ion mediating superexchange between two Os NN's does not sit at an inversion center and SO interactions are large for the 5$d$ shell. Hence, one should anticipate that the DM interaction is large, which our theoretical calculations confirm. We have demonstrated that the single ion anisotropy along with DM interaction and to a less extent the symmetric anisotropic exchange interaction, clearly has a role in the spin-wave gap of NaOsO$_3$. The contribution of the DM and $\Gamma$ interaction as the microscopic mechanism behind a spin-wave gap is seldom discussed, and NaOsO$_3$ is a unique material in this sense.

While our results for the magnon modes faithfully reproduce the dispersion reported on the basis of RIXS experiments\,\cite{Calder_2017}, we also point out that in the regime of observed magnons, the optical and acoustic modes are sometimes close in energy and in order to identify which mode is observed, it is important to theoretically analyse the scattering amplitudes. Furthermore, we find that a full account of all interaction parameters are needed in order to reproduce experimental observations of this complex compound. 

Finally, we note that it was argued in Ref.~\cite{Calder_2017} that NaOsO$_3$ is a system on the boundary between local-moment and itinerant magnetism, based on the fact that the SOC is very strong with a spin gap of $\unit[58]{meV}$ around the $\Gamma$ point. Our results can explain experimental magnon spectra by assuming an effective on-site Coulomb repulsion of $U_{\rm eff}= U-J_H =\unit[0.4]{eV}$ which drives the system out of the purely itinerant regime.

{\it Acknowledgements.} 
The authors acknowledge financial support from Knut and Alice Wallenberg Foundation through Grant No. 2018.0060. A.D. acknowledges financial support from the Swedish Research Council (VR) through Grants No. 2015-04608, No. 2016-05980, and No. 2019-05304. O.E. also acknowledges support from eSSENCE, SNIC, the Swedish Research Council (VR) and the ERC (synergy grant FASTCORR, project 854843). D.T. acknowledges support from the Swedish Research Council (VR) through Grant No. 2019-03666. E.S. acknowledges financial support from the Swedish Research Council (VR) through Grant No. 2017-03832. The work of Y.O.K. is supported by VR under the project No. 2019-03569.
Some of the computations were performed on resources provided by the Swedish National Infrastructure for Computing (SNIC) at the National Supercomputer Center (NSC), Linköping University, the PDC Centre for High Performance Computing (PDC-HPC), KTH, partially funded by the Swedish Research Council through grant agreement no. 2016-07213 and the High Performance Computing Center North (HPC2N), Umeå University.

\section{Supplementary Information}

\begin{table}[h]
 \caption{Calculated values of $J_{1}$, $D_{1}$ and $\Gamma_{1}$, for LDA+U approximation. Only values of $U$ and $J_{H}$ (in eV) are shown that produce a magnetic moment per Os atom close to the experimental value. All values of couplings are in meV.}
\vspace{5pt}
 \centering
 \begin{ruledtabular}
 \begin{tabular}{lcccc}
    \hline
      $U$ & $J_{H}$ & $J_{1}$ & $D_{1}$ & $\Gamma_{1}$   \\
      $3.5$  & $0.6$ &$-14.42$ & $ -1.67$ & $0.54$ \\
      $4.0$  & $0.6$ &$-16.87$ & $-1.67$ & $0.44$ \\
    
 \end{tabular}
 \end{ruledtabular}
    \label{tab:LDA_vs_LSDA}
\end{table}

\begin{figure}[h]
\includegraphics[width=8.5cm]{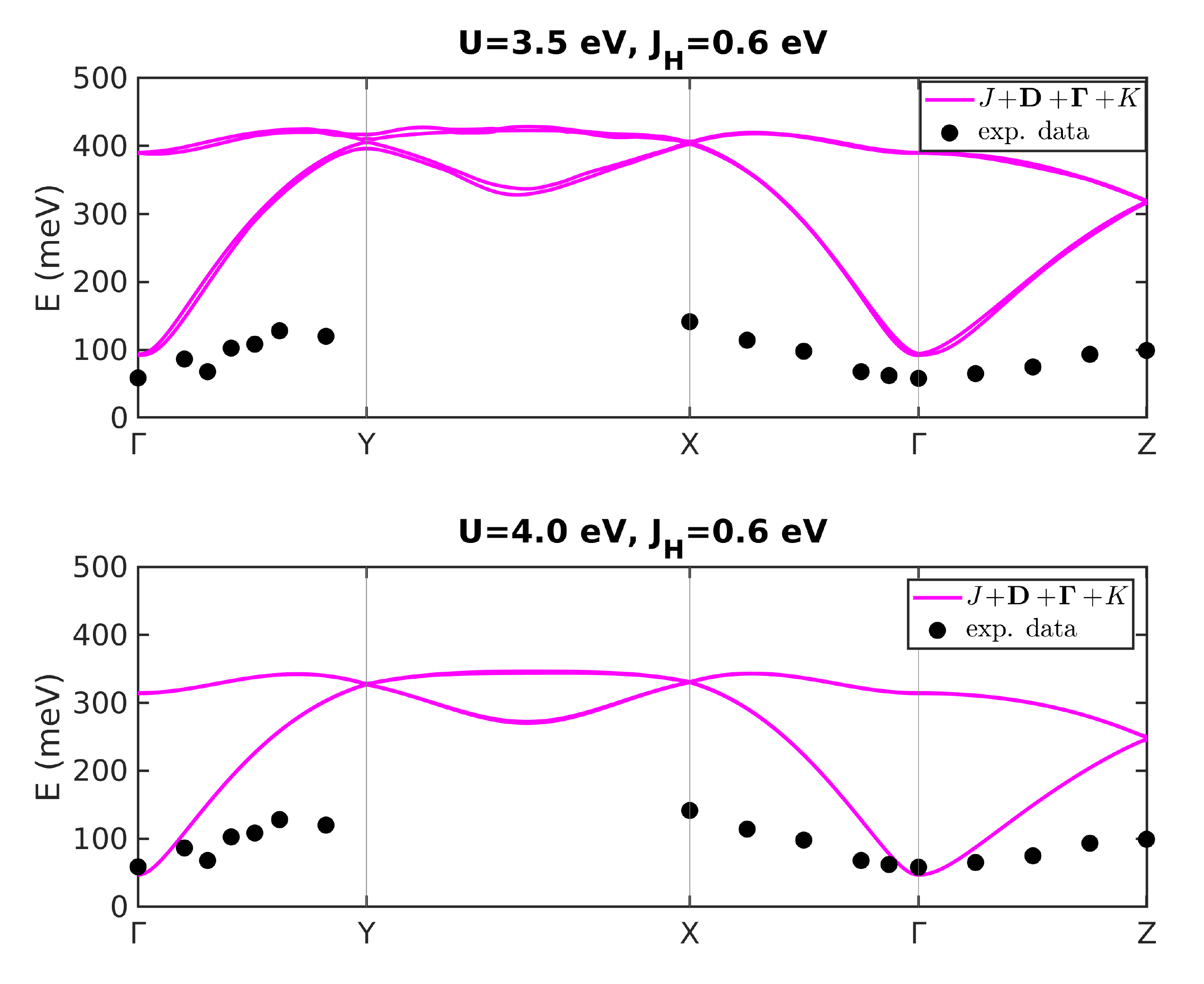}
\caption{
Adiabatic magnon spectrum of NaOsO$_3$ for LDA+U approximation. Only U=$3.5$ eV, $J_{H}=0.6$ eV and  U=$4.0$ eV, $J_{H}=0.6$ eV are shown as they produce a magnetic moment per atom close to the experimental one. Experimental data provided by RIXS measurements are shown in black dots \cite{Calder_2017}.
}
\label{fig_ams}
\end{figure}

\bibliographystyle{prb.bst}

\end{document}